\documentclass[12pt, a4paper]{article}
            
\usepackage{a4}
\usepackage{amsmath}
\usepackage{amssymb}
\usepackage[all]{xy}  
\usepackage{multirow}
\usepackage{cite}

\usepackage{hyperref}
\hypersetup{pdftitle={Computational Tools for Cohomology of Toric Varieties}, 
      pdfauthor={Ralph Blumenhagen, Benjamin Jurke, Thorsten Rahn}, 
      pdfsubject={Algebraic Geometry, Mathematical Methods in Physics},
      pdfstartview={FitH}, pdfpagelayout={TwoColumnRight},
      bookmarksopen, bookmarksnumbered, bookmarksopenlevel=2}

\newcommand{\cohomCalg}{\href{http://wwwth.mppmu.mpg.de/members/blumenha/cohomcalg/}{\text{\fontfamily{put}\bfseries\footnotesize\selectfont cohomCalg}}}
\newcommand{\cohomCalgKoszul}{{\text{\fontfamily{put}\bfseries\footnotesize\selectfont \href{http://wwwth.mppmu.mpg.de/members/blumenha/cohomcalg/}{cohomCalg} Koszul}~extension}}

\newcommand{\beq}{\begin{equation}}  \newcommand{\eeq}{\end{equation}}
\newcommand{\bal}{\begin{aligned}}   \newcommand{\eal}{\end{aligned}}

\def\IC{\mathbb{C}}
\def\IP{\mathbb{P}}
\def\IZ{\mathbb{Z}}
\def\IN{\mathbb{N}}

\def\cN{\mathcal{N}}
\def\cO{\mathcal{O}}
\def\cS{\mathcal{S}}
\def\cQ{\mathcal{Q}}
\def\cI{\mathcal{I}}

\def\fh{\mathfrak{h}}
\def\fC{\mathfrak{C}}

\def\clap#1{\hbox to 0pt{\hss#1\hss}}
\def\mllap{\mathpalette\mathllapinternal}
\def\mrlap{\mathpalette\mathrlapinternal}
\def\mclap{\mathpalette\mathclapinternal}
\def\mathllapinternal#1#2{%
\llap{$\mathsurround=0pt#1{#2}$}}
\def\mathrlapinternal#1#2{%
\rlap{$\mathsurround=0pt#1{#2}$}}
\def\mathclapinternal#1#2{%
\clap{$\mathsurround=0pt#1{#2}$}}

\def\ce{\mathrel{\mathop:}=}  

\def\fto{\longrightarrow}
\def\injto{\lhook\joinrel\relbar\!\!\:\joinrel\rightarrow}
\def\surjto{\relbar\joinrel\twoheadrightarrow}

\def\uspc{${}^\big.$}   

\begin{document}

\baselineskip=14pt
\parskip 5pt plus 1pt 

\vspace*{-1.5cm}
\begin{flushright}    
  {\small
  MPP-2011-39}
\end{flushright}

\vspace{2cm}
\begin{center}        
  {\LARGE
  Computational Tools for Cohomology \\[0.3cm]
  of Toric Varieties
  }
\end{center}

\vspace{0.75cm}
\begin{center}        
Ralph Blumenhagen, Benjamin Jurke and Thorsten Rahn
\end{center}

\vspace{0.15cm}
\begin{center}        
  \emph{Max-Planck-Institut f\"ur Physik, F\"ohringer Ring 6, \\ 
        80805 M\"unchen, Germany}
\end{center} 

\vspace{2cm}


\begin{abstract}
In this review, novel non-standard techniques for the computation of cohomology classes on toric varieties are summarized. After an introduction of the basic definitions and properties of toric geometry, we discuss a specific computational algorithm for the determination of the dimension of line-bundle valued cohomology groups on toric varieties. Applications to the computation of chiral massless matter spectra in string compactifications are discussed and, using the software package \cohomCalg, its utility is highlighted on a new target space dual pair of $(0,2)$ heterotic string models.
\end{abstract}

\clearpage

\newpage


\section{Introduction}
The computation of certain cohomology groups is a critical technical step in string model building, relevant for example in order to determine the (chiral) zero mode spectrum or parts of the effective four-dimensional theory, like the Yukawa coupling. Common methods often try to relate the computation at hand via a chain of isomorphisms back to known results in order to avoid most of the cumbersome computations from the ground up. Spectral sequences are the established technique to deal with such problems, but often end up to become laborious rather quickly. Having reasonable efficient algorithms to one's avail is therefore a vital requirement to make progress.

Supersymmetry in four dimensions, puts strong restrictions on the geometries admissible for string compactifications. In the absence of additional background fluxes (besides a gauge flux), this leads to the class of Calabi-Yau manifolds, where of particular  interest for ${\cal N}=1$ supersymmetry are Calabi-Yau threefold and fourfolds. Due to the Atiyah-Singer index theorem, chirality is realized by also turning on a non-trivial gauge background, which can be understood as the curvature of a non-trivial holomorphic vector bundle on the manifold. The majority of known Calabi-Yau manifolds is based on toric geometry. In particular, they are constructed as complete intersections of hypersurfaces in toric varieties. The vector bundle can then be described by different methods, where the three mostly used ones are: 
\begin{enumerate}
  \item the monad construction, which naturally arises in the $(0,2)$ gauged linear sigma model,
  \item the spectral cover construction, which gives stable holomorphic vector bundles with structure group $SU(n)$ on elliptically fibered Calabi-Yau threefolds,  
  \item the construction via extensions, which is the natural counter-part of brane recombinations.  
\end{enumerate}

All these three constructions have in common that they involve line bundles in one way or the other. For instance, the monad is defined via sequences of Whitney sums of line bundles, whereas the $n$-fold spectral cover is equipped in addition with a non-trivial line bundle on it, which via the Fourier-Mukai transform gives an $SU(n)$ vector bundle on the Calabi-Yau manifold. The basis starting point of every cohomology computation is therefore the knowledge of line bundle-valued cohomology classes on the ambient toric variety.

Using a simple yet powerful algorithm we  can compute the line bundle-valued cohomology dimensions $h^i(X;L_X)=\dim H^i(X;L_X)$ for any toric variety based on the information contained in the Stanley-Reisner ideal. The Koszul complex then allows to relate the cohomology on the toric variety to the cohomology of a hypersurface or complete intersection. The particular form of the algorithm also allows to easily deal with finite group actions on such geometries, i.e.~to consider orbifold spaces and twisted string states.

The review is organized as follows: In section~\ref{sec:ToricVarieties} some basics of toric geometry are introduced, including the Stanley-Reisner ideal and toric fans. Section~\ref{sec:Algorithm} introduces the computational algorithm for cohomology group dimensions of toric varieties that will be used throughout this review. Section~\ref{sec:Equivariance} shows how a finite group action and the resulting quotient space can be handled. In section~\ref{sec:KoszulComplex} the Koszul sequence is introduced, which allows to relate the ambient variety's cohomology to the cohomology of hypersurfaces and complete intersections. Monad bundle constructions and the Euler sequence are introduced in section~\ref{sec:MonadsEuler}. In section~\ref{sec:DualModel} we show an example of how to compute the data for a $(2,2)$ model that is dual to a $(0,2)$ model. The review closes in section~\ref{sec_outlook} with a brief outlook on potential further applications and developments.

\section{Toric Varieties}\label{sec:ToricVarieties}
One of the most important aspects of toric geometry is the ability to understand it in purely combinatorial terms, which is ideally suited to be handled by computers (see \cite{Fulton, Kreuzer:2006ax, Reffert:2007im,CoxLittleSchenk} for introductions into the subject). Toric geometry is also directly related to gauged linear $\sigma$-models (GLSMs) in physics \cite{Witten:1993yc}. On a more basic notion, a toric variety is a generalization of a projective space, which consists of a set of homogeneous coordinates $x_1,\dots,x_n$ as well as $R$ projective relations
\beq\label{eq_ToricProjectiveEquivalences}
  (x_1,\dots,x_n) \sim \Big(\lambda_r^{Q_1^{(r)}} x_1, \dots, \lambda_r^{Q_n^{(r)}} x_n \Big) 
  \qquad \text{for $\lambda_r\in\IC^\times$}.
\eeq
The $Q_i^{(r)}$ for $r=1,\dots,R$ and $i=1,\dots,n$ are GLSM charges, i.e.~the Abelian $U(1)$ charges in the associated GLSM, and corresponding to the projective weights. In direct comparison to projective spaces, toric varieties can be characterized as arising due to the usage of multiple projective relations instead of just a single one. The special case of a projective space therefore corresponds to $R=1$ in the above notation.

The homogeneous coordinates $x_i$ become  $\cN$=(2,2) chiral superfields in the GLSM picture and the Fayet-Iliopoulos parameters $\xi_r$ of the Abelian symmetries can be interpreted as the K\"ahler parameters of the geometric space. This parameter space of $\smash{\vec\xi=(\xi_1,\dots,\xi_R)}$ is then split into $R$-dimensional cones due to the vanishing of the D-terms associated to the GLSM. Within each cone the D-flatness condition can be solved and the cones correspond to the geometrical K\"ahler cones. Each such cone is often referred to as a geometric phase and can be fully characterized by a set of collections of coordinates
\beq
  \cS_\rho = \big\{ x_{\rho_1}, x_{\rho_2}, \dots, x_{\rho_{|\cS_\rho|}} \big\} \qquad \text{for $\rho = 1,\dots, N$}
\eeq
which are not allowed to vanish simultaneously. Note that such a collection is often written in product form, i.e.~the square-free monomial $x_{\rho_1} x_{\rho_2}\cdots x_{\rho_{|\cS_\rho|}}$ refers exactly to the same set. All those sets form the Stanley-Reisner ideal
\beq
  {\rm SR}(X) = \langle \cS_1,\dots, \cS_N \rangle,
\eeq
which can be equivalently used to uniquely specify a geometric phase. Note that the Stanley-Reisner ideal is Alexander-dual to the irrelevant ideal $B_\Sigma$ used in the mathematical literature.

Given the GLSM charges and the Stanley-Reisner ideal to identify the geometric phase, the toric variety $X$ of dimension $d=n-R$ can be described as the coset space
\beq
  X = (\IC^n - Z) \big/ {(\IC^\times)^R}.
\eeq
where $Z$ is the set of removed points specified by ${\rm SR}(X)$ via
\beq
  Z = \bigcup_{\rho=1}^N \big\{ x_{\rho_1} = x_{\rho_2} = \dots = x_{\rho_{|\cS_\rho|}} = 0 \big\}.
\eeq
This set $Z$ can be understood as the toric generalization of the removed origin in a projective space $\IC\IP^n= (\IC^{n+1}-\{0\})\big/\IC^\times$, as the Stanley-Reisner ideal for $\IC\IP^n$ is just the collection of all coordinates.

The combinatorial perspective on toric geometry mentioned at the start is formulated in terms of toric fans, cones and triangulations. In this language a geometric phase corresponds to a triangulation of a certain set of lattice vectors $\nu_i$ that span the fan $\Sigma_X$. The GLSM charges $\smash{Q_i^{(r)}}$ reappear in the form of $R$ linear relations
\beq\label{eq_LinearRelations}
  \sum_{i=1}^n Q_i^{(r)} \nu_i = 0 \qquad \text{for $r=1,\dots,R$}.
\eeq
By associating the lattice vectors $\nu_i$ to the homogeneous coordinates $x_i$, it becomes obvious that the linear relations \eqref{eq_LinearRelations} between the lattice vectors encode the projective equivalences \eqref{eq_ToricProjectiveEquivalences} between the homogeneous coordinates. In the language of fans the Stanley-Reisner ideal consists of all square-free monomials whose coordinates are not contained in any cone of the toric fan $\Sigma_X$.

\section[Dimensions of Line Bundle-valued Cohomology Groups]{Dimensions of Line Bundle-valued \\ Cohomology Groups}\label{sec:Algorithm}
Given a toric variety $X$ and a line bundle $L_X$, a frequent issue is to compute the $L_X$-valued cohomology group dimensions $h^i(X;L_X)$ for $i=0,\dots,\dim X$. After a couple of preliminary observations in \cite{Distler:1996tj,Blumenhagen:1997vt}, in \cite{Blumenhagen:2010pv} a complete novel algorithm for the determination of $h^i(X;L_X)$ was presented. This was subsequently proven in \cite{2010arXiv1006.0780J} and independently in \cite{Rahn:2010fm}.

The geometric input data for the computational algorithm presented below are the GLSM charges $\smash{Q_i^{(r)}}$ and the Stanley-Reisner ideal generators $\cS_1,\dots,\cS_N$. The basic idea of the algorithm is to count the number of monomials, where the total GLSM charge is equal to the divisor class of $D$, which is the divisor that specifies the line bundle $L_X = \cO_X(D)$. The form of those monomials is highly restricted by the Stanley-Reisner ideal, i.e.~the simpler the structure of ${\rm SR}(X)$, the easier the computation can be carried out. 

More precisely, negative integer exponents are only admissible for those coordinates that are contained in subsets of the Stanley-Reisner ideal generators. The most economic way is therefore to determine in a first step the set of square-free monomials $\cQ$ that arise from unions of the coordinates in any subset of ${\rm SR}(X)$. Each $\cQ$ gives a set of coordinates with negative exponents, and to each $\cQ$ there is an associated weighting factor $\fh_i(\cQ)$ that specifies to which cohomology group's dimension $h^i(X;\cO_X(D))$ the number of monomials $\cN_D(\cQ)$ with GLSM charge $D$ contributes. The cohomology group dimension formula can be summarized as
\beq
  \boxed{
  \dim H^i(X;\cO_X(D)) = \sum_{\cQ} \overbrace{\fh_i(\cQ)}^{\mclap{\text{multiplicity factor}}}\cdot\underbrace{\cN_D(\cQ)}_{\mclap{\text{number of monomials\;\;\;\qquad}}},\;{}}
\eeq
where the sum ranges over all square-free monomials that can be obtained from unions of Stanley-Reisner ideal generators. In the remainder of this section, both $\fh_i(\cQ)$ and $\cN_D(\cQ)$ will be properly defined.

\subsection*{Computation of multiplicity factors}
The multiplicity factors are defined by the dimensions of an intermediate relative homology. Let $[N]\ce \{1,\dots, N\}$ be a set of indices for the $N$ square-free monomials that generate the Stanley-Reisner ideal. Then let for each subset
\beq
  S_\rho \ce \{ \cS_{\rho_1},\dots,\cS_{\rho_k} \} \subset \{ \cS_1, \dots, \cS_N \}
\eeq
of generators $\cQ(S_\rho)$ be the square-free monomial that arises from the union of all coordinates in each generator $\cS_{\rho_i}$ of the subset.

The construction of the relative complex $\Gamma^\cQ$, from which $\fh_i(\cQ)$ is defined, goes as follows: From the full simplex on $[N]=\{1,\dots,N\}$ extract only those subsets $\rho\subset[N]$ with $\cQ(S_\rho)=\cQ$, i.e.~one considers all possible combinations of Stanley-Reisner ideal generators whose coordinates unify to the same square-free monomial $\cQ$. For some fixed $|\rho|=k$ this then defines the set of $(k-1)$-dimensional faces $F_{k-1}(\cQ)$ of the complex $\Gamma^\cQ$, i.e.~
\beq
  F_k(\cQ) \ce \left\{ \rho\subset [N] : \bal & |\rho| = k+1 \\ & \cQ(S_\rho) = \cQ \eal \right\}.
\eeq
Furthermore, let $\IC^{F_k(\cQ)}$ be the complex vector space with basis vectors $e_\rho$ for $\rho\in F_k(\cQ)$. The relative complex
\beq
  F_\bullet(\cQ): \qquad
  0 \fto F_{N-1}(\cQ)
  \stackrel{\phi_{N-1}}{\fto}
  \dots
  \stackrel{\phi_1}{\fto}
  F_0(\cQ)
  \stackrel{\phi_0}{\fto}
  F_{-1}(\cQ)
  \fto 0,
\eeq
where $F_{-1}(\cQ)\ce\{\emptyset\}$ is a face of dimension $-1$, is then specified by the chain mappings
\beq
  \bal
    \phi_k : {} & F_k(\cQ) \fto F_{k-1}(\cQ) \\
    {} & e_\rho \mapsto \sum_{s\in\rho} {\rm sign}(s,\rho)\, e_{\rho - \{s\}}.
  \eal
\eeq
A basis vector $e_{\rho - \{s\}}$ vanishes if $\rho$ with the element $s$ removed is not contained in $\Gamma^\cQ$. Furthermore, the signum is defined by ${\rm sign}(s,\rho) \ce (-1)^{\ell-1}$ when $s$ is the $\ell$th element of $\rho\subset[N]=\{1,\dots,N\}$ when written in increasing order.

For a given square-free monomial $\cQ$ then define the relative complex relabeling
\beq
  \fC_i(\cQ) \ce F_{|\cQ|-i}(\cQ)
\eeq
while leaving the mappings unchanged. The homology group dimensions
\beq
  \fh_i(\cQ) \ce \dim H_i(\fC_\bullet(\cQ))
\eeq
of the relabeled complex then provide the multiplicity factors that determine to which cohomology group $H^i(X;\cO_X(D))$ the monomials associated to $\cQ$ contribute. It should be emphasized that the $\fh_i(\cQ)$ only depend on the geometry (the Stanley-Reisner ideal) of the toric variety $X$ and not on the line bundle $\cO_X(D)$, i.e.~the multiplicity factors only have to be computed once for each geometry.

\subsection*{Counting monomials}
After computing the multiplicity factors $\fh_i(\cQ)$ it remains to count the number of relevant monomials. This second part of the algorithm depends on the GLSM charges of the homogeneous coordinates $x_i$ and the specific line bundle $\cO_X(D)$. Let $\cQ$ again be a square-free monomial. In order to simplify the notation, let $I=(i_1,\dots,i_k,\dots,i_n)$ be an index relabeling such that the product of the first $k$ coordinates gives $\cQ = x_{i_1}\cdots x_{i_k}$. Then one considers monomials of the form
\beq\label{eq_MonomialPrototype}
  \bal
    R^\cQ(x_1,\dots,x_n) &{} \ce (x_{i_1})^{-1-a} (x_{i_2})^{-1-b} \cdots (x_{i_k})^{-1-c} (x_{i_{k+1}})^d \cdots (x_{i_n})^e \\
    &{} = \frac{T(x_{i_{k+1}},\dots,x_{i_n})}{x_{i_1}\cdots x_{i_k} \cdot W(x_{i_1},\dots, x_{i_k})},
  \eal
\eeq
where $T$ and $W$ are monomials (not necessarily square-free) as well as exponents $a,b,c,d,e\in\IN \cup \{0\}$. One obviously finds the coordinates of the square-free monomial $\cQ$ in the denominator, whereas their complements are in the numerator. Based on the particular form of the relevant monomials define
\beq
  \cN_D(\cQ) \ce \dim \big\{ R^\cQ : \deg_{\rm GLSM}(R^\cQ) = D \big\},
\eeq
which counts the number of relevant monomials that have the same GLSM degree as the divisor $D$ that specified the line bundle $L_X=\cO_X(D)$.

\subsection*{A step by step example: del~Pezzo-1 surface}
\begin{table}[ht]
  \centering
  \begin{tabular}{r@{\,$=$\,(\,}r@{,\;\;}r@{\,)\;\;}|c|cc|c}
    \multicolumn{3}{c|}{vertices of the} & coords & \multicolumn{2}{c|}{GLSM charges} & {divisor class}${}^\big.$ \\
    \multicolumn{3}{c|}{polyhedron / fan} & & $Q^1$ & $Q^2$ & \\
    \hline\hline
    $\nu_1$ & $-1$ & $-1$ & $x_1$ & 1 & 0 & $H$\uspc \\
    $\nu_2$ &  1   &  0   & $x_2$ & 1 & 0 & $H$  \\
    $\nu_3$ &  0   &  1   & $x_3$ & 1 & 1 & $H+X$ \\
    $\nu_4$ &  0   & $-1$ & $x_4$ & 0 & 1 & $X$
  \end{tabular}
  \\[5mm] intersection form: ${}\quad HX - X^2$
  \\[3mm] ${\rm SR}(dP_1) = \langle x_1 x_2 ,\; x_3 x_4 \rangle = \langle \cS_1, \cS_2 \rangle$
  \caption{\small Toric data for the del Pezzo-1 surface}
  \label{tab:dPoneSurface}
\end{table}

In order to show the working algorithm in detail, we consider the del~Pezzo-1 surface. Its toric data is summarized in table~\ref{tab:dPoneSurface} for the reader's convenience. The two Stanley-Reisner ideal generators yield four possible combinations that become relevant in the computation, namely
\beq
  \cQ = 1, \;\; x_1 x_2,\;\; x_3x_4,\;\; x_1x_2x_3x_4.
\eeq
The computation of the multiplicity factors for those square-free monomials leads to
\beq
  \bal
  & \fC_0(1) = \big\{ \{ \emptyset \} \big\},\;\; \fC_1(x_1 x_2) = \big\{ \{\cS_1\} \big\}, \;\; \fC_1(x_3 x_4) = \big\{ \{ \cS_2 \} \big\}, \\
	& \fC_2(x_1x_2x_3x_4) = \big\{ \{\cS_1,\cS_2\} \big\}
	\eal
\eeq
and all other spaces $\fC_i(\cQ)$ vanishing. After computing the homology, this leads to the following contributions of the monomials \eqref{eq_MonomialPrototype} to the cohomology groups.
\beq
	\bal
		H^0(dP_1; \cO(m,n)):& \qquad T(x_1, x_2, x_3, x_4)\, ,\\
		H^1(dP_1; \cO(m,n)):& \qquad \frac{T(x_3,x_4)}{x_1 x_2\cdot W(x_1,x_2)}, \qquad \frac{T(x_1,x_2)}{x_3 x_4 \cdot W(x_3,x_4)}\, ,\\
		H^2(dP_1; \cO(m,n)):& \qquad \frac{1}{x_1 x_2 x_3 x_4 \cdot W(x_1,x_2,x_3,x_4)}\, .  
	\eal 
\eeq
Consider computing $h^\bullet(dP_1;\cO(-1,-2))$. Since all GLSM charges are positive, there is no contribution to $h^0$. Likewise, the denominator monomial of the $h^2$ contribution already has the GLSM charge $(3,2)$, which ``overshoots'' the target values and therefore also gives no contribution. $\smash{\deg_{\rm GLSM}(\frac{1}{x_1 x_2}) = (-2,0)}$ is no good either, but $\deg_{\rm GLSM}(\frac{1}{x_3x_4}) = (-1,-2)$ fits perfectly, such that there is a sole contribution
\beq
  \frac{1}{x_3x_4} \quad\leadsto\quad h^\bullet(dP_1;\cO(-1,-2)) = (0,1,0).
\eeq

All the aforementioned steps involved in the computation of the cohomology have been conveniently implemented in a high-performance cross-platform package called \cohomCalg\ \cite{cohomCalg:Implementation}.

\section[Equivariant Cohomology for Finite Group Actions]{Equivariant Cohomology \\ for Finite Group Actions}\label{sec:Equivariance}
Due to the explicit form of the relevant monomials that are counted by the algorithm, one can consider a rather simple generalization that also takes the action of finite groups into account \cite{Cvetic:2010ky, Blumenhagen:2010ed}. In orientifold and orbifold settings the internal part of the space-time is usually specified by a discrete symmetry acting on the ``upstairs'' geometry. This then induces a corresponding splitting of the cohomology groups 
\beq
  H^i(X) = H^i_{\rm inv}(X) \oplus H^i_{\text{non-inv}}(X)
\eeq
as the generating $p$-cycles can be either invariant or non-invariant under the symmetry. It is also necessary to specify the induced action on the bundle defined on the upstairs geometry. 

A so-called equivariant structure uplifts the action on the base geometry to the bundle and preserves the group structure. In fact, for a generic group $G$ each group element $g$ induces an involution mapping $g:X\fto X$ on the base geometry and has a corresponding uplift $\phi_g:V\fto V$ that has to be compatible with the bundle structure. This makes the diagram
\beq
  \parbox{1cm}{\xymatrix{ V \ar@{-->}[r]^{\phi_g} \ar@{->>}[d]_{\pi} & V \ar@{->>}[d]^{\pi} \\
	           X \ar[r]^{g} & X}} \qquad\leadsto\quad g\circ\pi = \pi\circ\phi_g
\eeq
commutative and the $G$-structure $V$ is called an equivariant structure, if it preserves the group structure, i.e.~if $\phi_g\circ\phi_h = \phi_{gh}$ holds such that the mapping $g\mapsto \phi_g$ is a group homomorphism.

The choice of an equivariant structure provides the means how the finite group acts on the relevant monomials \eqref{eq_MonomialPrototype} counted by the algorithm. For a given line bundle $\cO_X(D)$ one then has to check for all monomials whether or not they are invariant under the induced action. Consider for example the bundle $\cO(-6)$ on $\IC\IP^2$ and the $\IZ_3$ action
\beq\label{eq_Z3action}
  g_1:(x_1,x_2,x_3)\mapsto (\alpha x_1, \alpha^2 x_2, x_3) \qquad \text{for $\alpha\ce\sqrt[3]{1}={\rm e}^{\frac{2\pi{\rm i}}{3}}$}
\eeq
on the base coordinates. The same action is used for the monomials and thus defines the equivariant structure. The relevant monomials for the algorithm then pick up the following values from the involution:
\beq\label{eq_CP2algorithmCounting}
  \underbrace{
  \bal
	  \underbrace{\frac{1}{u_1^4u_2u_3}}_{g_1\to 1}, \quad
		\underbrace{\frac{1}{u_1u_2^4u_3}}_{g_1\to 1}, \quad
		\underbrace{\frac{1}{u_1u_2u_3^4}}_{g_1\to 1}, \quad
		\underbrace{\frac{1}{u_1^3u_2^2u_3}}_{g_1\to\alpha}, \quad
		\underbrace{\frac{1}{u_1^3u_2u_3^2}}_{g_1\to\alpha^2}, \\
		\underbrace{\frac{1}{u_1^2u_2^3u_3}}_{g_1\to\alpha^2}, \quad
		\underbrace{\frac{1}{u_1u_2^3u_3^2}}_{g_1\to\alpha}, \quad
		\underbrace{\frac{1}{u_1^2u_2u_3^3}}_{g_1\to\alpha}, \quad
		\underbrace{\frac{1}{u_1u_2^2u_3^3}}_{g_1\to\alpha^2}, \quad
		\underbrace{\frac{1}{u_1^2u_2^2u_3^2}}_{g_1\to 1},
	\eal}_{\displaystyle 	h^2(\IC\IP^2;\cO(-6))=(4_{\text{inv}},3_\alpha,3_{\alpha^2})},
\eeq
such that $h^\bullet_{\rm inv}(\IC\IP^2;\cO(-6))=(0,0,4)$ follows. This gives the cohomology of the quotient space $\IC\IP^2/\IZ_3$ as defined by the action in \eqref{eq_Z3action}.

This powerfull generalization of the algorithm allows for instance to compute the untwisted matter spectrum in heterotic orbifold models or (parts of) the instanton zero mode spectrum for Euclidean D-brane instantons in Type II orientifold models (see \cite{Blumenhagen:2010ja} for concrete applications).

\section{The Koszul Complex}\label{sec:KoszulComplex}
In most string theory applications, the geometries of interest are not toric varieties by themselves, but rather defined as subspaces thereof. These are defined as complete intersections of hypersurfaces of certain degrees. In order to relate the cohomology of the toric variety $X$ to the cohomology of a subspace, the Koszul sequence is used. 

To make this review self-contained and because it has been implemented in the \cohomCalgKoszul\  package, let us briefly describe how this works. Let $D\subset X$ be an irreducible hypersurface and $0\not=\sigma\in H^0(X;\cO(D))$ be a global non-zero section of $\cO_X(D)$, such that $Z(\sigma)\cong D$. This induces a mapping $\cO_X\fto \cO_X(D)$ and its dual $\cO_X(-D) \injto \cO_X$, the latter of which can be shown to be injective. Given an effective divisor
\beq
  D \ce \sum_i a_i H_i \subset X
\eeq
where all $a_i \ge 0$, there is a short exact sequence
\beq\label{eq_PlainKoszulSequence}
  0 \fto \cO_X(-D) \injto \cO_X \surjto \cO_D \fto 0,
\eeq
called the Koszul sequence. Here $\cO_D$ is the quotient of the sheaf $\cO_X$ of holomorphic functions on $X$ by all holomorphic functions vanishing at least to order $a_i$ along the irreducible hypersurface $H_i\subset X$. This allows to treat $\cO_D$ as the structure sheaf on the divisor $D$, which effectively identifies the sheaf cohomology $H^i(X;\cO_D)$ with $H^i(D;\cO_D)$. A proper definition of the involved mappings, which become quite laborious to work out explicitly, can be found in \cite{GriffithsHarris}. In addition of the plain Koszul sequence \eqref{eq_PlainKoszulSequence}, there is also a twisted variant
\beq\label{eq_TwistedKoszulSequence}
  0 \fto \cO_X(T-D) \injto \cO_X(T) \surjto \cO_D(T) \fto 0
\eeq
that is obtained by tensoring \eqref{eq_PlainKoszulSequence} with the line bundle $\cO_X(T)$. The induced long exact cohomology sequence
\begin{small}
  \beq
    \parbox{1cm}{\xymatrix{ 
        0 \ar[r] & H^0(X;\cO_X(T-D)) \ar[r] & H^0(X;\cO_X(T)) \ar[r] & H^0(D;\cO_D(T)) \ar`[rd]`[l]`[dlll]`[d][dll] & \\
                 & H^1(X;\cO_X(T-D)) \ar[r] & H^1(X;\cO_X(T)) \ar[r] & H^1(D;\cO_D(T)) \ar`[rd]`[l]`[dlll]`[d][dll] & \\
                 & H^2(X;\cO_X(T-D)) \ar[r] & H^2(X;\cO_X(T)) \ar[r] & H^2(D;\cO_D(T)) \ar[r]                       & \dots}}
  \eeq
\end{small}
then allows to relate the cohomology of the toric variety $X$ directly to the cohomology of the hypersurface.

Given a more generic case of several (mutually transverse) hypersurfaces $\{S_1,\dots,S_l\}$ one can compute the cohomology on the complete intersection via the generalized Koszul sequence
\begin{small}
    \beq\label{eq_generalizedkoszulwithdiviso}
      \parbox{1cm}{\xymatrix{ 
          & 
            \hspace{-1.3cm}0 \fto \smash{\cO_X\bigg({\textstyle-\sum\limits_{j=1}^l S_{j}+D}\bigg)} \ar[r] & 
            \ldots \ar[r] & 
            \smash{\bigoplus\limits_{\mclap{i_1<i_2}} \cO_X\left(-S_{i_1}-S_{i_2}+D\right)} \ar`[rd]`[l]`[dlll]`[d][dll] & 
            \\ & 
            \smash{\bigoplus\limits_{i_1} \cO_X(-S_{i_1}+D)} \ar[r] & 
            \cO_X(D)\big. \ar[r] & 
            \cO_S(D) \fto 0\,. &  
        }}^\big._\big.
    \eeq
\end{small}
In contrast to the hypersurface sequence, this is no longer a short exact sequence and hence does not  give rise to a long exact sequence in cohomology. One way to proceed is via the technique of spectral sequences, which inductively allows one to compute the wanted cohomology classes on the complete intersection. However, for our implementation, we decided to take a different approach. We break down this long  sequence \eqref{eq_generalizedkoszulwithdiviso} into several short exact sequences using several auxiliary sheaves $\cI_k$:
\beq\label{eq_generalizedkoszulsplitted}
    \bal
        0\fto \cO_X\bigg({\textstyle -\sum\limits_{j=1}^l S_{j}+D}\bigg) \injto
        & {} \bigoplus_{\mclap{i_1<\ldots<i_{l-1}}} \cO_X\bigg({\textstyle -\sum\limits_{j=1}^{l-1}S_{i_j}+D}\bigg) \surjto \cI_1 \fto 0 \\
        0\fto\cI_1 \injto 
        & {} \bigoplus_{\mclap{i_1<\ldots<i_{l-2}}} \cO_X\bigg({\textstyle-\sum\limits_{j=1}^{l-2}S_{i_j}+D}\bigg) \surjto \cI_2 \fto 0 \\
        & {}\quad \vdots \\
        0\fto\cI_{l-2} \injto 
        & {} \bigoplus_{i_1}\cO_X\left(-S_{i_1}+D\right) \surjto \cI_{l-1} \fto 0 \\
        0\fto\cI_{l-1} \injto
        & {} \cO_X(D) \surjto \cO_S(D) \fto 0
    \eal
\eeq
The individually induced long exact sequences of cohomology can then be used for the step-wise computation of $H^\bullet(S;\cO_S(D))$, which is the cohomology on the complete intersection $\smash{S=\bigcap_{i=1}^l S_i}$.

\section{Monad Construction of Vector Bundles}\label{sec:MonadsEuler}
Before we come to a concrete application in heterotic string model building, let us present the construction of holomorphic vector bundles via a so-called monad. Such a structure directly arises in the $(0,2)$ GLSM description and can be regarded as a generalization of the tangent bundle of a complete intersection in a toric variety. 

Given the GLSM charges defined in \eqref{eq_ToricProjectiveEquivalences}, the tangent bundle can be defined as the quotient $T_S={\rm Ker}(f)/{\rm Im}(g)$ of the sequence
\beq\label{eq_generaleuler}
  0 \fto \overbrace{\cO_S^{\oplus R}}^{\mclap{\substack{\text{one $\cO_S$ for each} \\ \text{Picard generator}}}} \stackrel{g}{\injto}
  \underbrace{\bigoplus_{i=1}^n \cO_S(Q_i)}_{\mclap{\substack{\text{one bundle with the GLSM} \\ \text{charges for each coordinate}}}} \stackrel{f}{\surjto}
  \overbrace{\bigoplus_{j=1}^l \cO_S(S_j)}^{\mclap{\substack{\text{one bundle with the degree} \\ \text{for each hypersurface}}}} \fto 0
\eeq
where the individual line bundles are restricted to the complete intersection $\smash{S=\bigcap_{i=1}^l S_i}$. The rank of the resulting vector bundle is given by ${\rm rk}(T)=n-l-R$. Using the methods presented so far, it is clear that they allow to compute the dimensions of the cohomology classes $h^i(S;T_S)$, where the initial input data for the set of long exact sequences are the line bundle valued cohomology classes on the ambient toric variety.

The $(0,2)$ GLSM generalizes this in the sense that the bundle the left-moving world-sheet fermions couple to is not any longer the tangent bundle of the Calabi-Yau, but a more general holomorphic (stable) vector bundle $V$, which is analogously defined via a sequence of Whitney sums of line bundles
\beq\label{eq_general monad}
  0 \fto \cO_S^{\oplus R_V} \stackrel{g}{\injto} 
  \bigoplus_{a=1}^\delta \cO_S(N_a) \stackrel{g}{\surjto}
  \bigoplus_{l=1}^\lambda \cO_S(M_l) \fto 0.
\eeq
The rank is ${\rm rk}(V)=\delta-\lambda-R_V$. The charges $N_a$ and $M_l$ have to satisfy the anomaly cancellation conditions 
\beq\label{eq_anomaly cancellation 1}
  \bal
    &\sum_a N_a^{(\alpha)} = \sum_l M_l^{(\alpha)},  \qquad \forall~\alpha\,,\\
    &\sum_l M_l^{(\alpha)} M_l^{(\beta)} - \sum_a N_a^{(\alpha)} N_a^{(\beta)} =  
     \sum_j S_j^{(\alpha)} S_j^{(\beta)} - \sum_i  Q_i^{(\alpha)} Q_i^{(\beta)}, \quad \forall~\alpha,\beta,
  \eal
\eeq
where $1\le \alpha,\beta\le R$ denote the components corresponding to the $U(1)$ actions in the GLSM. The most delicate issue for such constructions is the proof of $\mu$-stability. However, it should be clear that besides that the monad construction provides a large set of heterotic $(0,2)$ backgrounds and that the methods described so far are indeed taylor-made for the determination of the zero mode spectrum, which is given by the dimensions of vector bundle valued cohomology classes $h^i(S;\Lambda^k V )$.

\section{A (2,2) Model Dual to a (0,2) Model}\label{sec:DualModel}
Now let us show all this for  concrete heterotic $(0,2)$ models, for which we first recall a couple of issues. The theory is naturally equipped with an $E_8\times E_8$ gauge theory. One of these $E_8$'s may be taken to be invisible to the real world and hence only one $E_8$ remains. The holomorphic vector bundle now is endowed with a certain structure group $G$ which breaks this $E_8$ down to some GUT group. The remaining GUT group is then simply the commutant of $G$ in $E_8$. Depending on what kind of GUT group we are interested in, we may choose the structure group $G$ to be either $SU(3)$, $SU(4)$ or $SU(5)$ breaking $E_8$ down to $E_6$, $SO(10)$ or $SU(5)$ respectively.

In order to obtain the number of zero modes in different representations of the GUT group we have to calculate the cohomology classes of bundles involving the holomorphic vector bundle \cite{WittenCohomology}. The precise correlation of vector bundle cohomology and zero modes for all three GUT groups are given in table \ref{table_GUT group representations via cohomology} (for a nice review on the particle spectrum of heterotic theories see for instance \cite{HeteoticParticleSpectrum}).
\begin{table}
  \small
	\newcommand{\Xoplus}{\mllap{\oplus\,}}
  \begin{tabular}{l|cccccc}
    \# zero modes & &&&&&\\
    in reps of $H$&  1 &   $h^1_S(V)$ & $h^1_S(V^\ast)$ &  $h^1_S(\Lambda^2 V)$ &  $h^1_S(\Lambda^2 V^\ast)$ & $h^1_S(V\otimes V^\ast)$  \\
    \hline\hline
    \qquad\;\;\,\parbox{1cm}{$E_8$\uspc \\ ${}\,\downarrow$} & \multicolumn{6}{|c}{\parbox{1cm}{248\uspc \\ ${}\;\downarrow$}}  \\ 
    $SU(3) \times E_6$     &  $(1,78)$ & $\Xoplus(3,27)$ & $\Xoplus(\overline{3},\overline{27})$  & & & $\Xoplus(8,1)$\uspc \\
    $SU(4) \times SO(10)$  & $(1,45)$ & $\Xoplus(4,16)$ & $\Xoplus(\overline{4},\overline{16})$ & $\Xoplus(6,10)$ & & $\Xoplus(15,1)$\\
    $SU(5) \times SU(5)$   &  $(1,24)$ & $\Xoplus(5,\overline{10})$ & $\Xoplus(\overline{5}, 10)$ & $\Xoplus(10,5)$ & $\Xoplus(\overline{10},\overline{5})$ & $\Xoplus (24,1)$\\
  \end{tabular}
  \caption{\small Correlation between zero modes in representations of the GUT group $H$}
  \label{table_GUT group representations via cohomology}
\end{table}
The moduli appearing in such a framework are given by possible deformations of the Calabi-Yau manifold, which are counted by the Hodge numbers 
\beq
  h^{2,1}(S) \text{ and } h^{1,1}(S)
\eeq
and by possible deformations of the bundle, i.e. the bundle moduli, which are counted by the dimension of the cohomology of the endomorphism bundle End$(V)$ of $V$. Furthermore one can show that
\beq
  H^1(S;\text{End}(V)) \cong H^1(S;V^*\otimes V)\,,
\eeq
which simplifies its determination. In case of the standard embedding, the vector bundle is simply the tangent bundle and hence has $SU(3)$ structure and gauge group $E_6$. Many vector bundles can be constructed using monads, by defining the vector bundle to be the cohomology of the complex \eqref{eq_general monad}. Using only this complex, it is possible to construct  bundles with the structure groups shown  in table \ref{table_GUT group representations via cohomology} and hence computing all these cohomologies  simply boils down to the computation of line bundle cohomology on the complete intersection. This on the other hand can be related, using the Koszul sequence \eqref{eq_generalizedkoszulwithdiviso}, to the cohomology of line bundles on the ambient toric variety.

In the following we give an example of a pair of heterotic models which are related by a so-called target space duality \cite{Distler:1995bc,Chiang:1997kt,Blumenhagen:1997vt} and were derived in \cite{BRTargetSpaceDuality}. The first of those will be a $(2,2)$ model $(M_a,V_a)=(M_a,T_{M_a})$ while the second one, referred to as $(M_b,V_b)$, is of type $(0,2)$ equipped with an $SU(3)$-bundle which is assumed to be stable. 

Let us start with an example in which we can already see most of the structure but which is not too involved. Consider
\beq
  V_{1,1,1,1,2,2,2}[3,4,3]\surjto \IP^6_{1,1,1,1,2,2,2}[3,4,3] \,.
\eeq
Since this configuration is singular we have to resolve it by introducing a new coordinate. This yields the smooth configuration shown in table~\ref{tab_TwoTwoModel}, leading to the following monad for the tangent bundle:
\begin{table}[t]
  \centering
  \begin{tabular}{cccccccc|ccc}
    \multicolumn{8}{c|}{coordinate} & \multicolumn{3}{c}{hypers.${}^\big.$} \\
    \multicolumn{8}{c|}{GLSM charges} & \multicolumn{3}{c}{degrees} \\
    \hline\hline
    0 & 0 & 0 & 0 & 1 & 1 & 1 & 1    & 1 & 1 & 2\uspc \\
    1 & 1 & 1 & 1 & 2 & 2 & 2 & 0    & 3 & 3 & 4
  \end{tabular}
  \caption{\small Toric data for the smooth $(2,2)$ model 3-fold geometry $M_a$}
  \label{tab_TwoTwoModel}
\end{table}
\begin{table}[t]
  \centering
  \begin{tabular}{cccccccccc|cccc}
    \multicolumn{10}{c|}{coordinate} & \multicolumn{4}{c}{hypersurf.${}^\big.$} \\
    \multicolumn{10}{c|}{GLSM charges} & \multicolumn{4}{c}{degrees} \\
    \hline\hline
    0 & 0 & 0 & 0 & 0 & 0 & 0 & 0 & 1 & 1    & 0 & 0 & 1 & 1\uspc \\
    0 & 0 & 0 & 0 & 1 & 1 & 1 & 1 & 0 & 0    & 1 & 1 & 1 & 1 \\
    1 & 1 & 1 & 1 & 2 & 2 & 2 & 0 & 0 & 0    & 3 & 3 & 2 & 2 \\
  \end{tabular}
  \caption{\small Toric data for the dual $(0,2)$ model 3-fold geometry $M_b$}
  \label{tab_ZeroTwoModel}
\end{table}
\beq
  \parbox{2cm}{\xymatrix{\mllap{0\fto{}}\cO_{M_a}^{\oplus 2} \ar@{^{(}->}[r] & \cO_{M_a}(0,1)^{\oplus4}\oplus \cO_{M_a}(1,2)^{\oplus3} \oplus \cO_{M_a}(1,0) \ar@{->>}[d] \\
  & \cO_{M_a}(1,3)^{\oplus2}\oplus\cO_{M_a}(2,4) \mrlap{{}\fto 0,}}}
\eeq
where the Koszul sequence \eqref{eq_TwistedKoszulSequence} has to be applied as well. Using {\cohomCalgKoszul} we can obtain the number of zero modes of the chiral spectrum in this model as well as the dimension of the moduli space:
\beq\label{eq_match chiral spectrum example 1}
  \bal
    h^\bullet_{M_a}(V_a) &{} = ( 0, 68 , 2,0),\\
    h^{1,1}_{M_a}+h^{2,1}_{M_a}+h^1_{M_a}({\rm End}(V_a)) &{}=  2 + 68 + 140 = 210,
  \eal
\eeq
where the reader should keep in mind that in this case $V_a=T_{M_a}$ is just the tangent bundle. The dual $(0,2)$ model geometry can then be determined to be the data in table~\ref{tab_ZeroTwoModel}, and its monad is specified by the sequence
\beq
  \parbox{2cm}{\xymatrix{\mllap{0 \fto{}} \cO_{M_b}^{\oplus2} \ar@{^{(}->}[d] \\
  \mclap{\cO_{M_b}(0,0,1)^{\oplus4}\oplus\cO_{M_b}(0,1,2)\oplus\cO_{M_b}(1,0,0)\oplus\cO_{M_b}(0,2,4)\oplus\cO_{M_b}(0,1,0)} \ar@{->>}[d] \\
  \cO_{M_b}(0,1,3)^{\oplus2} \oplus \cO_{M_b}(1,2,4)  \mrlap{{}\fto 0}}}
\eeq
This configuration satisfy the conditions (\ref{eq_anomaly cancellation 1}) and we obtain the following topological data:
\beq\label{eq_match chiral spectrum 2}
  \bal
    h_{M_b}^\bullet(V_b) &{}= ( 0, 68 , 2, 0)\,,\\
    h_{M_b}^{1,1}+h_{M_b}^{2,1}+h_{M_b}^1 (\text{End}(V_b)) &{}=  3 + 51 + 156 = 210\,.
  \eal
\eeq
Comparing to the data (\ref{eq_match chiral spectrum example 1}) we can see that the number of zero modes in the chiral spectrum does not change and even though the individual Hodge numbers as well as their sum are both different, the dimension of the full moduli space stays the same. 

This is a manifestation of a so far only very poorly understood perturbative (in $g_s$) target space duality in the configuration space of heterotic string compactifications with ${\cal N}=1$ supersymmetry in four dimensions.

\section{Outlook}\label{sec_outlook}
So far most implementations of computational methods in string model building were based on toric geometry \cite{PALP} and in particular on the combinatorial formulas of Batyrev and Borisov \cite{Borisov,Batyrev:258454,batyrev-1994}\footnote{Of course there also general software tools for algebraic geometry like \cite{Rambau:TOPCOM-ICMS:2002,M2, SAGE}.}. Clearly, these are very powerful but also have their limitations. First, they only apply in the $(2,2)$ case, where the vector bundle is identified with the tangent bundle. Second, for complete intersections the combinatorial formulas only hold for so-called nef-partitions which ensure that the corresponding polytopes representing the space are reflexive.

The computational tool reviewed in this article can also be applied to situations where other packages fail. As explained, the powerful algorithm for the determination of the dimensions of line-bundle valued cohomology classes is taylor-made for dealing also with general complete intersection and for $(0,2)$ models, where the vector bundle is defined via line bundles, e.g.~the monad construction or the spectral cover construction.

Of course, also the algorithm implementation {\cohomCalg} has its limitations. First, in situations where the number of Picard generators (projective relations, reflected by $h^{1,1}$) becomes large (about the order of ten), the computations become too involved and the program too time consuming. A second drawback is the exponential growth of the computing time with the number of Stanley-Reisner ideal generators, which at the moment takes several hours for about 40 generators. Third, if there are not enough zeros in the many intermediate long exact sequences, the result is not unique and one has to determine the kernel respectively image of maps by hand.

\subsection*{Acknowledgments}
The authors would like to thank Helmut Roschy for his contributions to the original work presented in this review.

\appendix


\bibliography{rev9}
\bibliographystyle{utphys}

\end{document}